\documentclass[11pt,a4paper]{article}

\usepackage{amsmath}
\usepackage{amsfonts}
\usepackage{amssymb}
\usepackage{graphicx,psfrag,color}
\usepackage{subfigure}
\usepackage{hyperref}

\usepackage{enumerate}

\usepackage[left=2cm,top=2.5cm,right=2cm,bottom=2cm]{geometry}

\begin{document}
%\noindent \textbf{Mod. Phys. Lett. A, Vol. XX, No. XX (XXXX) XXXXXXX}
\begin{center}
\Large{\textbf{Constraints on Dark Matter-Photon Coupling in the Presence of Time-Varying Dark Energy}} \\[0.2cm]
 
\large{\textbf{ Santosh Kumar Yadav\footnote{E-mail: sky91bbaulko@gmail.com}}}
\\[0.1cm]

\small{
\textit{ Department of Mathematics, BITS Pilani, Pilani Campus, Rajasthan-333031, India.}}

\end{center}

\vspace{.2cm}

\noindent \small{\textbf{\large{Abstract:}} In a recent work [Phys. Rev. D 98, 043521 (2018)], we have investigated a dark matter (DM)-photon coupling model in which the DM decays into photons in the presence of dark energy (DE) with constant equation of state (EoS) parameter. Here, we study an extension of the DM-photon coupling model by considering a time-varying EoS of DE via Chevalier-Polarski-Linder (CPL) parametrization. We derive observational constraints on the model parameters by using the data from cosmic microwave background (CMB), baryonic acoustic oscillations (BAO), the local value of Hubble constant from Hubble Space Telescope (HST), and large scale structure (LSS) information from the abundance of galaxy clusters, in four different combinations. We find that in the present DM-photon coupling scenario the mean values of $w_{\rm de0}$ are in quintessence region ($w_{\rm de0} > -1$) whereas they were in the phantom region ($w_{\rm de0} < -1$) in our previous study with all data combinations. The constraints on the DM-photon coupling parameter do not reflect any significant deviation from the previous results. Due to the decay of DM into photons, we obtain higher values of $H_0$, consistent with the local measurements, similar to our previous study. But, the time-varying DE leads to lower  values of $\sigma_8$ in the DM-photon coupling model with all data combinations, in comparison to the results in our previous study. Thus, allowing time-varying DE in the DM-photon coupling scenario is useful to alleviate the $H_0$ and $\sigma_8$ tensions.\\

\noindent \small{\textbf{{PACS:}} 95.35.+d, 95.36.+x, 98.80.Cq
\section{Introduction}
\label{sec:intro}

The major matter component, namely, dark matter (DM) is a mysterious component of the Universe whose precise nature is still an open question in modern cosmology. In the literature, many attempts have been made  via direct and indirect searches to know about the nature/properties of DM like mass, spin, parity, interaction cross-section etc. but very less is known concretely so far. See \cite{DM01,DM02} for a review about the evidences, candidates and methods of detection of DM. The phenomenon of decay of DM into species like dark radiation, photons, neutrinos etc. has been considered in the literature in different contexts and motivations. A review of decaying DM signals in gamma-rays, cosmic ray antimatter and neutrinos can be seen in \cite{DM03}. More intensively, the search for DM decay has been carried out using the IceCube telescope data \cite{IceCube}. Many theoretical/phenomenological studies have been caried out with DM decay models in order to look for some possible solutions to the problems associated with the standard $\Lambda$CDM cosmology. For instance, the evidence for DM-dark radiation interaction is reported in \cite{DDM06} where it has been found that this interaction allows to reconcile the $\sigma_8$ tension between Planck cosmic microwave background (CMB) and large scale structure (LSS) measurements. 
It has been observed in \cite{DM05,DM06} that the late-time decay of DM is helpful in reconciling some of the small-scale structure formation problems  associated with the standard $\Lambda$CDM cosmology. Also see \cite{DDM01,DDM02,DDM03,DDM04,DDM05,DDM05_1,DDM05_2}, where the interaction between DM and dark radiation has been investigated.

The decay of DM into photons (and photons + neutrinos) has been investigated  from cosmic-ray emission in \cite{Dgamma01,Dgamma02,Dgamma03}. 
An analytical and  numerical study of DM-photon intractions have been performed in \cite{DM07} where some consequences of DM-photon interaction on structure formation have  been explored. 
Recently, the constraints on DM-photon scattering-cross section in the early Universe have been obtained in \cite{DM08}. The upper bounds on the decay width of DM into different final states can be investigated by searching decaying DM. 
An upper limit on the DM-photon elastic scattering cross section  $\sigma_{{\rm DM}-\gamma} \lesssim 10^{-32} \, (m_{\rm DM}/{\rm GeV}) \, \rm cm^2$ has been derived in \cite{DM09}. 
An upper bound on elastic scattering cross section of DM-neutrino and DM-dark energy (DE) have been obtained  as $\sigma_{{\rm DM}-\nu} \lesssim 10^{-33} \, (m_{\rm DM}/{\rm GeV})\, {\rm cm^2} $ and
$\sigma_{\rm DM-DE} \lesssim 10^{-29} \, (m_{\rm DM}/{\rm GeV})\, {\rm cm^2} $ in \cite{DM10} and \cite{DM11}, respectively.

Recently, we (two more authors) have studied a DM-photon coupling model in \cite{DDM07}, where the constraints on decay rate of DM into photons, and possible consequences of the coupling scenario are investigated in the presence of DE with constant equation of state (EoS) parameter. In the work \cite{DDM07}, we have considered interaction between DM and photons where the DM decay into photons takes place throughout the cosmic history of the Universe, leading to non-conservation of the particle number densities for both the species. Note that this interaction is different from the DM-photon elastic scattering interaction considered in \cite{DM08, DM09, DM10,DM11}, where the particle number density remains conserved.  Here, we present a follow-up study of \cite{DDM07} by considering a time-varying EoS of DE  via Chevalier-Polarski-Linder (CPL) parametrization  \cite{CPL01,CPL02}. The main aim of this work is to investigate the possible changes/effects of time-varying DE on the results of the DM-photon coupling scenario obtained in our recent study \cite{DDM07}. We constrain this scenario by using recent data from CMB, baryonic acoustic oscillations (BAO), the local value of Hubble constant from Hubble Space Telescope (HST), and LSS information from the abundance of galaxy clusters, in four different combinations. 

The rest of this work is organized as follows. In the next Section \ref{model}, we describe the DM-photon coupling model and discuss the possible effects of variable EoS of DE on CMB temperature (TT) and matter power spectra. Section \ref{data} describes the observational data and methodology used in this work.  In Section \ref{results}, we derive observational constraints and discuss the results. We conclude the main findings of this study in Section \ref{remarks}. 
In what follows, a subindex 0 attached to any quantity denotes its present value and a prime over a quantity represents its derivative with respect to conformal time.

\section{DM-photon Coupling Model with Time-varying Dark Energy} 
\label{model}
In this section, we reproduce the DM-photon coupling scenario and the related perturbation equations considered in our recent study \cite{DDM07} while inducing the time-varying EoS of DE via CPL parametrization. We assume that, in the framework of Friedmann-Lema\^{i}tre-Robertson-Walker (FLRW) Universe, DM decays into photons via the  
the following background density equations:

\begin{align}
\label{ddm_fundo}
 \rho'_{\rm ddm} + 3 \frac{a'}{a} \rho_{\rm ddm} = - \frac{a'}{a} \Gamma_{\gamma} \rho_{\rm ddm}, 
\end{align}
\begin{align}
\label{gamma_fundo}
 \rho'_{\gamma} + 4 \frac{a'}{a} \rho_{\gamma} = \frac{a'}{a} \Gamma_{\gamma} \rho_{\rm ddm}, 
\end{align}
 where the dimensionless parameter $\Gamma_{\gamma}$ characterizes the DM-photons coupling. The quantities $\rho_{\rm ddm}$ and $\rho_{\gamma}$  denote  energy density of decaying DM and photons, respectively.
 
For the $i$th coupled component, the covariant conservation
equation allowing for energy-momentum transfer
gives $\nabla_{\mu} T^{\nu \mu}_{ i} = Q^{\nu}_i$. The requirement of the conservation of the total
energy-momentum, $\nabla_{\mu} T^{\nu \mu} = 0$, demands $\sum_i Q^{\nu}_i = 0$. Here, it is satisfied with $Q_{\rm ddm}= -\frac{a'}{a}\Gamma_\gamma\, \rho_{\rm ddm}$ and $Q_{\gamma}= \frac{a'}{a}\Gamma_\gamma\, \rho_{\rm ddm}$. Note that a non-conservation in the number density of DM particles results in non-conservation of the energy-momentum tensor of the DM particles. We assume $\Gamma_{\gamma}>0$ to have a  decaying DM throughout the  expansion history of the Universe. In gerenal, the DM decay rate is considered to be constant but it could also be time-varying as well. We define decay rate as follows: $\Gamma = \Gamma_\gamma \mathcal{H}/a$, where $\mathcal{H}$ is conformal Hubble parameter. Here, we extend our previous  analysis \cite{DDM07} by considering a time-varying EoS of DE via CPL parametrization \cite{CPL01,CPL02,CPL03,CPL04}, given by
\begin{align}\label{CPL}
w_{\rm de}(a) &= w_{\rm de 0} + w_{\rm de 1}(1-a),
\end{align}
where $w_{\rm de0}$ and $w_{\rm de1}$ are free parameters (constants) to be constrained by the observational data.
The evolution of density of decaying DM and photons can be easily found from eqs. \eqref{ddm_fundo} and \eqref{gamma_fundo} or more explicitly see eqs. (3) and (4) in \cite{DDM07}.

\subsection{Perturbation Equations}
In the present work, we consider the linear perturbations in synchronous gauge via the line element of the linearly perturbed FLRW metric:
\begin{align}
 ds^2 = - a^2d\tau^2 + a^2[(1- 2 \eta)\delta_{ij} + 2 \partial_i \partial_j E]dx^idx^j,
\end{align}
where $k^2 E = -2/h -3 \eta$, restricted to the scalar modes $h$ and $\eta$. Then, using $\nabla_{\mu} T^{\nu \mu}_{\rm i} = Q^{\nu}_i$, the continuity and Euler equations of the $i$th coupled fluid read as follows: 

\begin{eqnarray}
 \delta'_i + 3\mathcal{H} (c^2_{s,i} -w_i)\delta_i + 9\mathcal{H}^2 (1+w_i) (c^2_{s,i} - c^2_{a,i}) \frac{\theta_i}{k^2}  
 + (1+w_i)\theta_i -3(1+w_i)\eta' 
 + (1+w_i) \left(\frac{h'}{2} + 3\eta' \right)  \\ \nonumber
   = \frac{a}{\rho_i}(\delta Q_i - 
 Q_i \delta_i) + a \frac{Q_i}{\rho_i} \Big[3\mathcal{H} (c^2_{s,i} - c^2_{a,i}) \Big] \frac{\theta_i}{k^2},
\end{eqnarray}

\begin{eqnarray}
\theta'_i + \mathcal{H} (1 - 3c^2_{s,i})\theta_i - \frac{c^2_{s,i}}{(1+w_i)} k^2 \delta_i   
= \frac{a Q_i}{(1 + w_i)\rho_i} \Big[\theta_{\rm ddm} - (1+c^2_{s,i})\theta_i \Big],
\end{eqnarray}
where $c^2_{a,i}$, $c^2_{s,i}$ and $w_i$, respectively represent the
adiabatic sound speed, physical sound speed and EoS of the $i$th fluid in the rest frame (See \cite{lixin14} and references therein for such a methodology to describe the linear perturbations of the interaction between DM and DE). 

Further, by particularizing the fluid approximation equations to the DM and photon coupled system, the continuity and Euler equations for photons, respectively, read as follows:

\begin{eqnarray}
\label{delta_gamma}
\delta'_{\gamma} + \frac{4}{3}\theta_{\gamma} + \frac{2}{3}h' = a \Gamma_{\gamma} \mathcal{H} \frac{\rho_{\rm ddm}}{\rho_{\gamma}} (\delta_{\rm ddm} - \delta_{\gamma}),
\end{eqnarray}

\begin{eqnarray}
\label{theta_gamma}
 \theta'_{\gamma} - \frac{1}{4}k^2 (\delta_{\gamma} - 4\sigma_{\gamma}) - a n_e \sigma_T (\theta_b - \theta_{\gamma})  
 = \frac{3}{4} a \Gamma_{\gamma} \mathcal{H} \frac{\rho_{\rm ddm}}{\rho_{\gamma}} (\theta_{\rm ddm} - \frac{4}{3} \theta_{\gamma}),
\end{eqnarray}
where $\theta_b$ is the divergence of baryons fluid velocity, and  $a n_e \sigma_T (\theta_b - \theta_{\gamma})$ 
appears due to the collision 
 between photons and baryons before recombination. The momentum transfer is chosen in the rest frame of DM. 
The DM evolution is given by
\begin{eqnarray}
 \delta'_{\rm ddm} + \frac{h'}{2} = 0.
\end{eqnarray}
In the synchronous gauge, the Euler equation for DM reads as
\begin{eqnarray}
\theta_{\rm ddm} = 0.
\end{eqnarray} 

\subsection{Effects on Matter and CMB TT Power Spectra}
As discussed in our previous study \cite{DDM07}, the matter power spectrum, CMB anisotropies, CMB
spectral distortions, luminosity distance etc. can be affected in various ways due to the non-conservation of the photon number density resulting from the decay of DM into photons.

\begin{figure*}[!ht]
\centering
 \includegraphics[width=6.5cm]{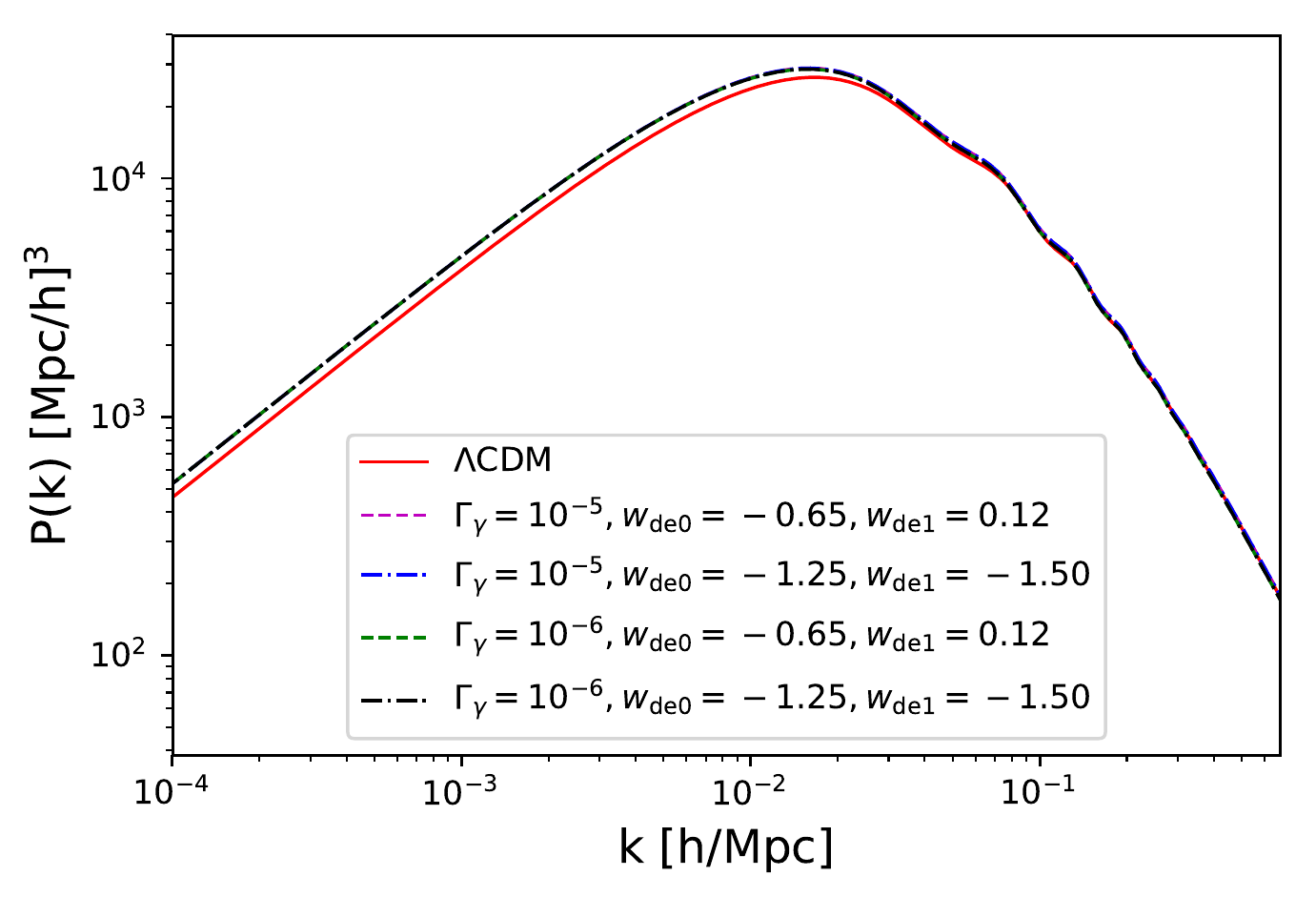}
 \includegraphics[width=6.5cm]{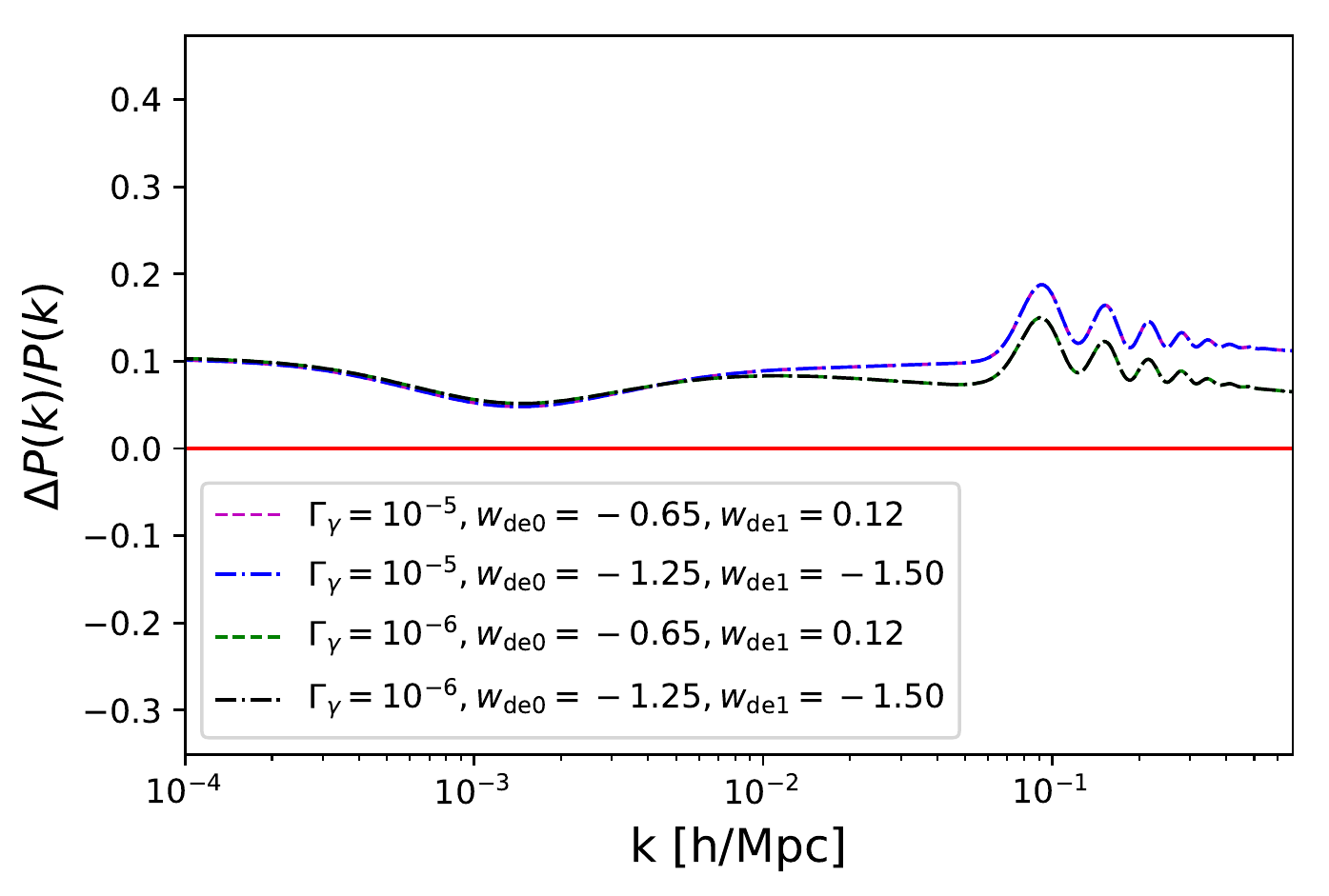}
\caption{\label{matter} {The matter power spectrum and its relative deviations from  standard $\Lambda$CDM model for some values of $\Gamma_\gamma$, $w_{\rm de0}$ and  $w_{\rm de1}$ mentioned in legend whereas the other related parameters are kept to their respective mean value from Table \ref{Table_M1}.}}
\end{figure*}

\begin{figure*}[!ht]
\centering
 \includegraphics[width=6.5cm]{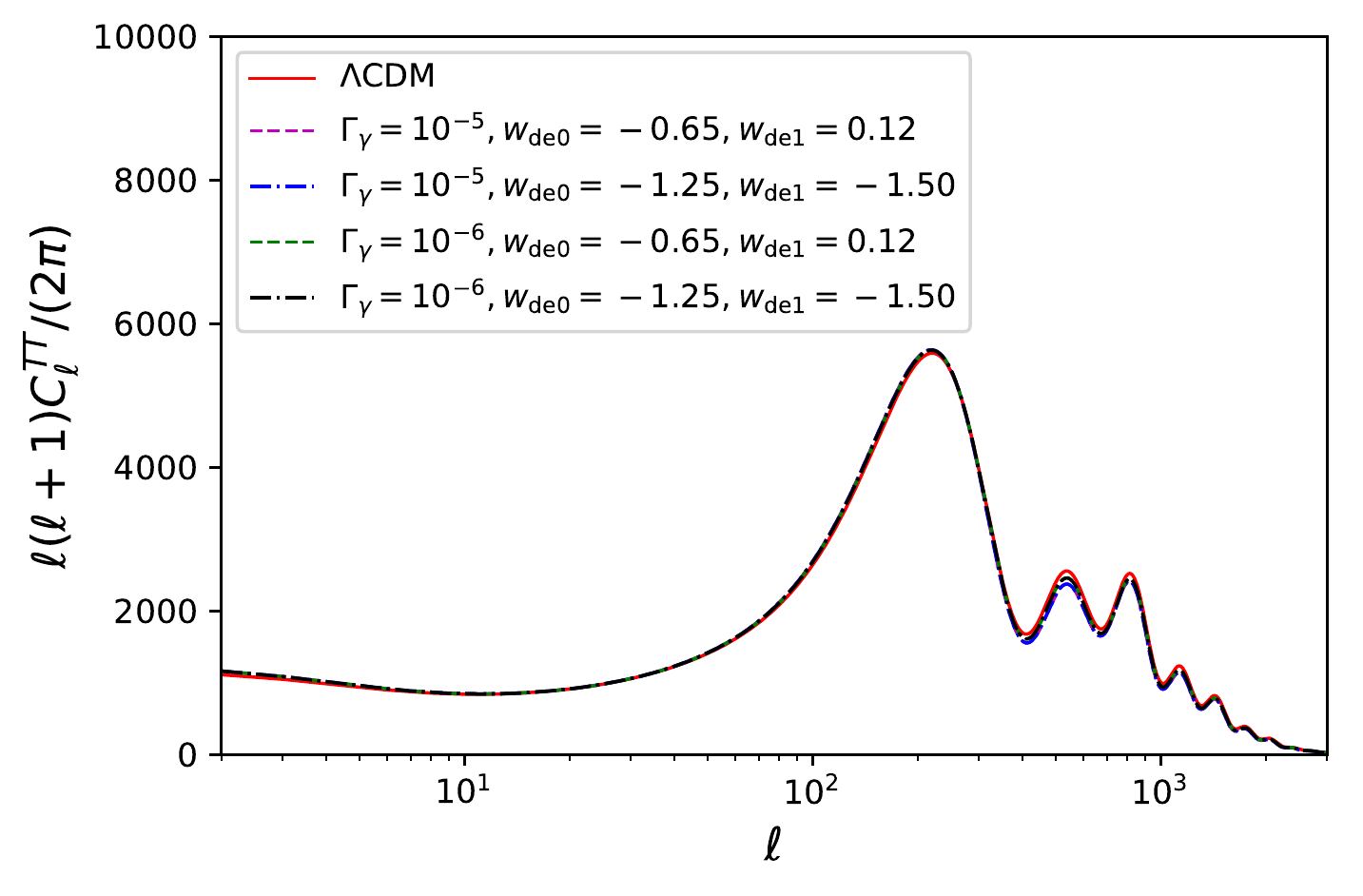}
 \includegraphics[width=6.5cm]{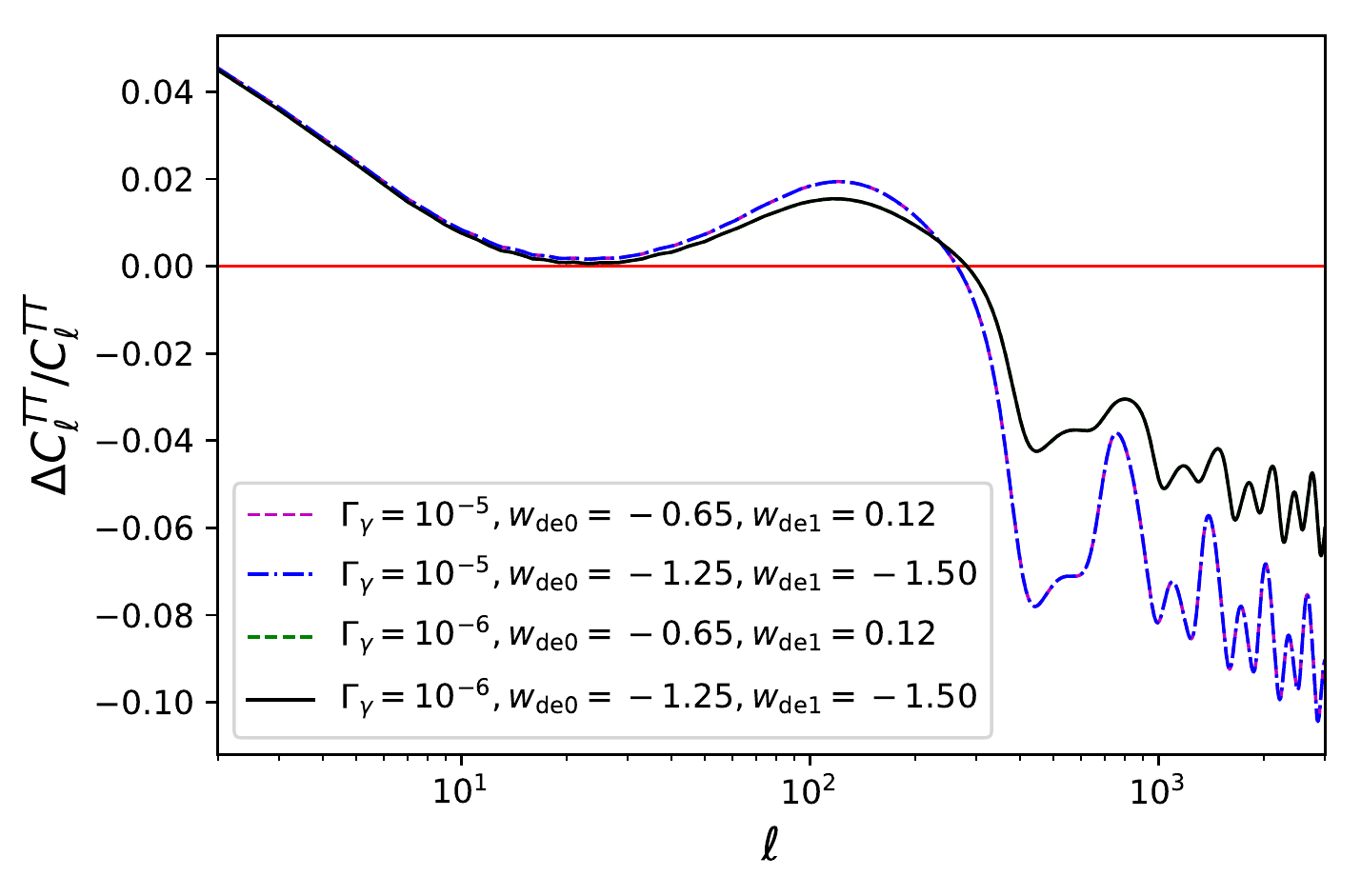}
\caption{\label{cmb} {The CMB TT power spectrum and its relative deviations from  standard $\Lambda$CDM model for some values of $\Gamma_\gamma$, $w_{\rm de0}$ and  $w_{\rm de1}$ mentioned in legend whereas the other related parameters are kept to their respective mean value from Table \ref{Table_M1}.}}
\end{figure*}  

Figure \ref{matter} and Fig. \ref{cmb} respectively show the matter and CMB TT power spectra with their relative deviations from  standard $\Lambda$CDM model for some values of the parameters $\Gamma_\gamma$, $w_{\rm de0}$ and  $w_{\rm de1}$, as mentioned in legends whereas the other related parameters are kept to their respective mean value from Table \ref{Table_M1}. We notice that the two spectra deviate considerably from the $\Lambda$CDM model due to change   in $\Gamma_\gamma$ as we observed in our previous study. On the other hand, no deviations are observed due to change in the EoS parameters of DE. Thus, the time-varying DE does not affect the matter and CMB power spectra on the top of the DM-photon coupling scenario.

In any general modification of $\Lambda$CDM cosmology (within DE models), it  is expected that the main effects of CMB anisotropies occur on the amplitude of the late time integrated Sachs-Wolfe (ISW) effect, manifested at large angular scales. This effect depends on the duration of the DE domination, i.e., on the time of equality of matter and DE density. So, different behaviors of $w(z)$ (in this case from CPL model), quintessence or phantom behavior must have opposites effects in $l < 100$. So, the values of $w_{\rm de0}$ and $w_{\rm de1}$ can be fixed in such a way that DE will show quintessence or phantom behavior at late time (for instance, for $z < 2$).  The constraints on  total matter density  ($\Omega_{\rm m}$)  control the amplitude of peaks, especially the second and third peaks. It can be seen  from Table \ref{Table_M1}, that the changes on $\Omega_{m}$ are minimal, so the amplitude corrections will also be minimal. 
Also, changes in the expansion of the Universe, at late time from CPL free parameters and early times from $\Gamma_{\gamma}$ will contribute to the corrections on the amplitude of all peaks and shifts on the spectrum due to the modification in the angular diameter distance at decoupling (depend on the expansion history of the DM-photon interaction model after decoupling). The magnitudes of  corrections are proportional to the possible deviations from the values, $\Gamma_{\gamma}  = 0, w_{\rm de0} = -1$, and $w_{\rm de1} = 0$ , compared with minimal $\Lambda$CDM model. Note that in our work, $\Gamma_{\gamma} \ll 1$.  The couping parameter $\Gamma_{\gamma}$ will contribute at small scales because it changes the density of photons at $z \gg 1$.

%%%%%%%%%%%%%%%%%%%%%%%%%%%%%%%%%%%%%%%%%%
%\section{Discussion}

\section{Data, Methodology  and Model Parameters} 
\label{data}

The observational data sets used to derive constraints on various model parameters are described briefly as follows. \\

\noindent\textbf{CMB}: Planck cosmic microwave background data consisting of high-$l$ (TT), low-$l$ polarisation and Planck lensing survey from Planck-2015 \cite{Planck2015}.\\

\noindent\textbf{BAO}: Four baryon acoustic oscillations observations from the  Six  Degree  Field  Galaxy  Survey  (6dFGS) at  $z_{\rm eff} = 0.106$ \cite{bao1}, 
the  Main  Galaxy  Sample  of  Data  Release 7  of  Sloan  Digital  Sky  Survey  (SDSS-MGS) at $z_{\rm eff} = 0.15$ \cite{bao2}, 
the  LOWZ  and  CMASS  galaxy  samples  of Data Release 11 of  the Baryon  Oscillation  Spectroscopic  Survey  (BOSS) LOWZ  and  BOSS-CMASS at  $z_{\rm eff} = 0.32$ and $z_{\rm eff} = 0.57$, respectively \cite{bao3}. These BAO observations are summarized in Table I of \cite{baotot}.
\\

\noindent\textbf{HST}: The local value of Hubble constant, $H_0=73.24 \pm 1.74$  km s${}^{-1}$ Mpc${}^{-1}$ by Hubble Space Telescope (HST) \cite{riess}. \\

\noindent\textbf{LSS}: Three large scale structure (LSS) observations, including the  measurements from the Sunyaev-Zeldovich (SZ) effect cluster mass function: $\sigma_8\left(\frac{\Omega_m}{0.27}\right)^{0.30} = 0.782 \pm 0.010$ \cite{LSS1};  weak gravitational lensing data from Canada-France-Hawaii  Telescope Lensing Survey (CFHTLenS): $\sigma_8\left(\frac{\Omega_m}{0.27}\right)^{0.46} = 0.774 \pm 0.040$ \cite{LSS2}; and the weak gravitational lensing shear power spectrum constraints: $\sigma_8\left(\frac{\Omega_m}{0.30}\right)^{0.50} = 0.651 \pm 0.058$ from the Kilo Degree Survey (KiDS-450) \cite{LSS3}.

The underlying model is implemented in publicly available CLASS \cite{class} code, and the parameter inference is done by using the Monte Python \cite{monte} code which is embedded with Metropolis Hastings algorithm and interfaced with CLASS code to obtain correlated Markov Chain Monte Carlo (MCMC) samples. The uniform priors used in this work are mentioned in Table \ref{tab:priors}.  The correlated MCMC samples are obtained with four different data combinations: CMB + BAO, CMB + BAO + HST, CMB + BAO + LSS, and CMB + BAO + HST + LSS. The convergence of the Monte Carlo Markov Chains is checked by Gelman-Rubin criteria \cite{Gelman_Rubin} which requires $1-R< 0.03$ for all the model parameters, in general. GetDist Python package \cite{antonygetdist} is used to analyse the obtained MCMC samples.

In the present work, we have considered variable EoS of DE in CPL form with the motivation to investigate its possible effects and 
deviations from our previous study \cite{DDM07}. We have also included neutrino mass scheme following normal hierarchy with a minimum sum of neutrino mass 0.06 eV. We have considered $N_{\rm eff}$, effective number of relativistic species as a free parameter. 
Finally, the base parameters set of the underlying model is:
 \begin{align}\nonumber
  \{100\omega_{\rm b}, \, \omega_{\rm cdm}, \, 100\theta_{s}, \, \ln10^{10}A_{s}, \, n_s, \, \tau_{\rm reio}, 
  \, w_{\rm de0}, \, w_{\rm de1},\, \sum m_{\nu}, \, N_{\rm{eff}}, \, \Gamma_{\gamma} \},
 \end{align} 
where the first six parameters pertain to the standard $\Lambda$CDM model \cite{p13}.

\begin{table}
\caption{Uniform priors on model parameters used in the present work.} \label{tab:priors}
\begin{center}
\begin{tabular}{|c|c|}
\hline
Parameter & Prior\\
\hline
$100 \omega_{\rm b}$ & [1.8,	2.4]\\
$\omega_{\rm cdm}$ & [0.001,	0.99] \\
$100\theta_s$ & [0.5,	10.0] \\
$\ln[10^{10}A_{s }]$ & [2.7,	4.0]\\
$n_s$ & [0.9,	1.1] \\
$\tau_{\rm reio}$  & [0.01,	0.9] \\
$w_{\rm de0}$  & [-2.0,	0.5] \\
$w_{\rm de1}$ &  [-1.5,	1.5] \\
$\sum m_{\nu}$ & [0.06, 1.0]\\
$N_{\rm eff}$ & [1.0,  4.0]\\
$\Gamma_{\gamma}$ & [0, 0.0001]\\
\hline
\end{tabular}
\end{center}
\end{table} 

\section{Results and Discussion}
\label{results}
The observational constraints on baseline parameters and some derived parameters of the underlying model are shown in Table \ref{Table_M1} with four data combinations: CMB + BAO, CMB + BAO + HST, CMB + BAO + LSS and CMB + BAO + HST + LSS (joint analysis). The first six parameters are well consistent with the standard $\Lambda$CDM cosmology.
With all  data combinations, the mean values of $w_{\rm de0}$ indicate  quintessence behaviour ($w_{\rm de0}>-1$) of DE. See the one-dimensional marginalized distribution of $w_{\rm de0}$ in left panel of Fig. \ref{Gamma}, where the vertical dotted line corresponds to $w_{\rm de0} = -1$ (EoS of the DE given by cosmological constant). On the other hand, in our previous work \cite{DDM07} with constant EoS of DE, the mean values of $w_{\rm de0}$ were  in the phantom region ($w_{\rm de0}<-1$) with all data combinations. The DM-photon coupling parameter $\Gamma_\gamma$ is approximately of the order $10^{-5}$ (upper 95\% CL) with all data combinations under consideration (see the one-dimensional marginalized distribution of $\Gamma_\gamma$ in the right panel of Fig. \ref{Gamma}). These constraints on $\Gamma_\gamma$ are  similar to those obtained in our previous work \cite{DDM07} where a constant EoS of DE was assumed. Thus, the time-varying EoS of DE does not have any significant effect on the DM-photon coupling parameter $\Gamma_\gamma$.
 
\begin{table*}[] 
\caption{\label{Table_M1} {Constraints (68\%  and 95\% CL) on the free parameters and some derived model parameters with four different data combinations are displayed. The parameters $H_0$  and  $\sum m_{\nu} $ are measured in the units of km s${}^{-1}$ Mpc${}^{-1}$ and eV,   respectively. The $\chi^2_{\rm min}$ values of the fit are also shown in last row.}}
\resizebox{\textwidth}{!}{%
\begin{tabular} { c c c c c}  \hline \hline 
 Parameter &  CMB + BAO   & CMB + BAO + HST & CMB + BAO + LSS & CMB + BAO + HST + LSS \\
\hline

$10^{2}\omega_{\rm b }$ &    $2.22^{+ 0.15+0.19}_{-0.08-0.22}$   & $2.32^{+0.08+0.08}_{-0.03-0.12}  $ &
$2.25^{+ 0.14+0.16}_{-0.06-0.21}$  & $2.31^{+0.08+0.09}_{-0.03 -0.13}   $\\[1ex]

$\omega_{\rm cdm }  $ &  $0.121^{+0.013+0.019}_{-0.009-0.021}            $ & $0.131^{+0.007+0.013}_{-0.006-0.014}   $ &$0.121^{+0.012+0.018}_{-0.008-0.021}$& $0.127^{+0.007+0.011}_{-0.006-0.013}$ \\[1ex]

$100 \theta_{s }$  &  $1.0415^{+0.0009+0.0020}_{-0.0010-0.0018}$  &$1.0407^{+0.0007+0.0014}_{-0.0007-0.0013}$   &
$1.0414^{+0.0008+0.0021}_{-0.0010-0.0018}$  &  $1.0409^{ +0.0007+0.0014}_{-0.0007-0.0014} $ \\[1ex]
 
$\ln10^{10}A_{s }$   &  $3.095^{+0.037+0.079}_{-0.043-0.074}$      &$3.102^{+0.039+0.079}_{-0.039-0.076}$    &                    $3.115^{+0.043+0.083}_{-0.043-0.082}$    &      $3.112^{+ 0.042+0.085}_{-0.042-0.085}$ \\[1ex]

$n_{s }$       &    $0.974^{+0.013+0.025}_{-0.013-0.024}$   &$0.975^{+0.012+0.024}_{-0.012-0.024}$        &    $0.976^{+0.012+0.025}_{-0.013-0.023}$         &  $0.977^{+0.012+0.026}_{-0.013-0.023} $\\[1ex]

$\tau_{\rm reio }$   &  $0.080^{+0.017+0.038}_{-0.020-0.035}   $ &$0.080^{+0.019+0.038}_{-0.019-0.038}   $  & $0.092^{+0.019+0.038}_{-0.019-0.038}    $& $0.092^{+ 0.020+0.040}_{-0.020-0.036} $    \\[1ex]

$w_{\rm de0} $ & $-0.76^{+0.24+0.37}_{-0.16-0.43}$   &    $-0.89^{+0.18+0.32}_{-0.16-0.33}$  &    $-0.86^{+0.24+0.38}_{-0.16-0.43}$ &    $-0.93^{+0.19+0.30}_{-0.14-0.33}$  \\[1ex]

$w_{\rm de1} $ & $-0.85^{+0.22+1.00}_{-0.62-0.68}$   &    $-0.62^{+0.42+0.98}_{-0.59-0.88}$  &    $-0.88^{+0.18+0.96}_{-0.61-0.66}$ &    $-0.80^{+0.26+0.92}_{-0.65-0.72}$  \\[1ex]

$\sum m_{\nu} [95\%\, \rm CL]$  & $<0.39 $   &  $< 0.52$   &  $<0.86   $ &   $ <0.89 $\\ [1ex]

$N_{\rm{eff}}$   & $3.29^{+0.39+0.77}_{-0.39-0.81}$  &     $3.60^{+0.32+0.65}_{-0.32-0.61}$  &   $3.40^{+0.43+0.85}_{-0.43-0.89}$   & $3.62^{+ 0.34+0.68}_{-0.34-0.65}$\\[1ex]

$\Gamma_{\gamma } [95\%\, \rm CL]$      & $<2.7\times10^{-5}$ & $<5.1\times 10^{-6}$   & $<2.2\times10^{-5}$     &   $<7.7 \times10^{-6}$\\ [1ex]

\hline

$\Omega_{\rm{m} }$&   $0.320^{+0.023+0.043}_{-0.023-0.045} $  &     $0.302^{+ 0.016+0.031}_{-0.016-0.030}$  &  $0.308^{+0.024+0.043}_{-0.021-0.046}            $  &    $0.299^{+ 0.016+0.031}_{-0.016-0.031}$   \\[1ex]
  
  $H_{\rm 0}$ & $67.4^{+3.9 +8.0}_{-3.9-8.0} $
 &    $72.2^{+ 1.6+3.2}_{-1.6-3.0} $   & $69.8^{+ 4.1+8.0}_{-4.1-8.0}               $  & $72.5^{+1.5+2.9}_{-1.5-2.9}$   \\[1ex]
 
 $\sigma_8$ & $0.799^{+0.020+0.052}_{-0.026-0.045}$   &   $0.816^{+ 0.022+0.042}_{-0.020-0.047} $ & $ 0.761^{+ 0.017+0.040}_{-0.021-0.037} $         &$0.767^{+0.014+0.031}_{-0.016-0.028} $\\[1ex]
  
 $r_{\rm{drag} }$ & $146.4^{+4.8+13}_{-7.9-11} $   &     $140.3^{+2.9+7.5}_{-4.1-6.1}$  & $144.8^{+ 4.5+14.0}_{-7.5-11.0}$
   &  $141.0^{+ 2.8+7.6}_{-4.3-6.3} $\\[1ex]
  
 \hline
  
 $\chi^2_{\rm min}/2   $ & $5640.95$  &  $5641.80$  & $5648.78  $ & $5649.02 $\\[1ex]  
  \hline \hline
\end{tabular}}
\end{table*}
 
\begin{figure*}[!ht] \centering
\includegraphics[width=7.5cm]{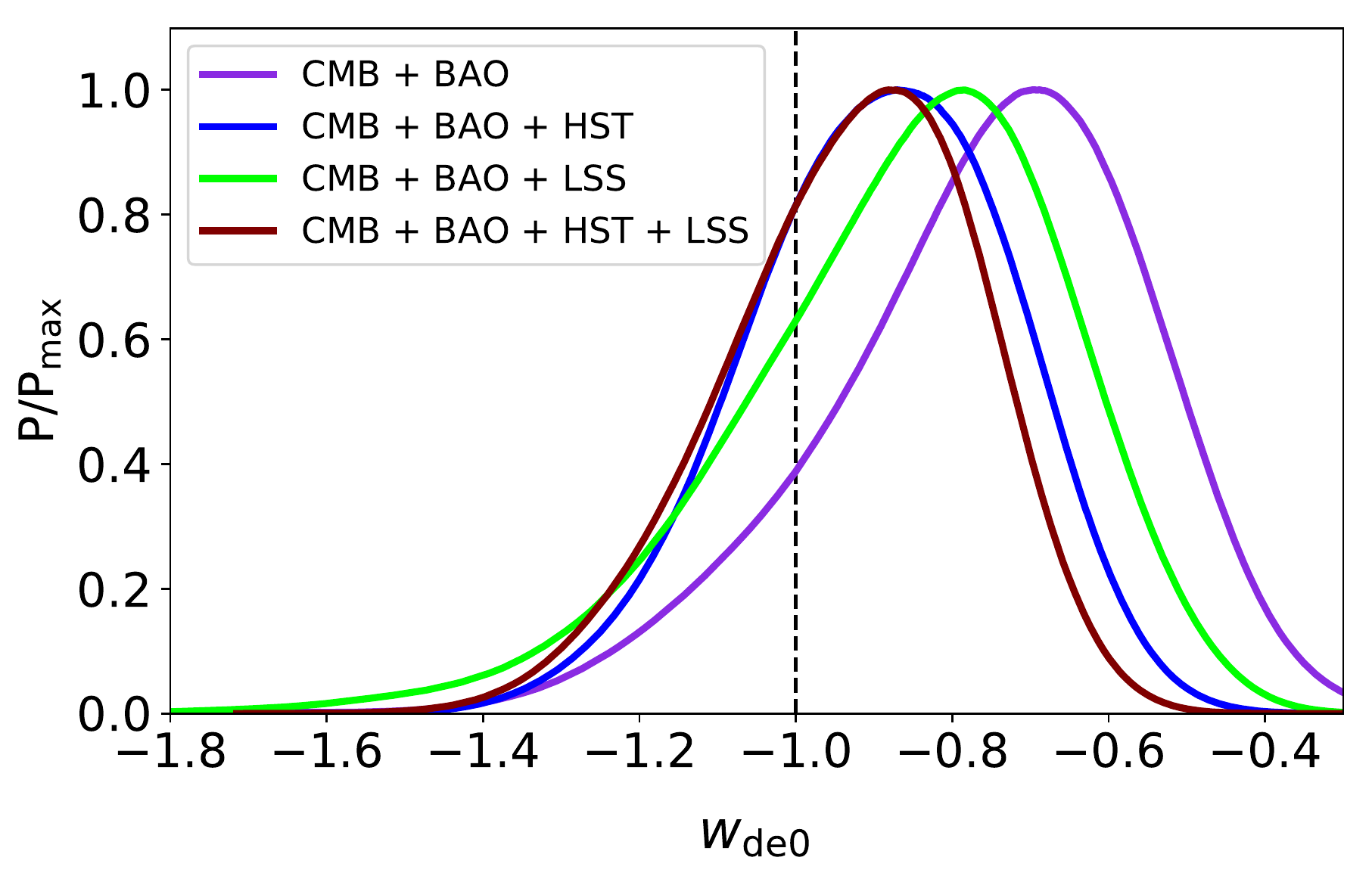}
 \includegraphics[width=7.5cm]{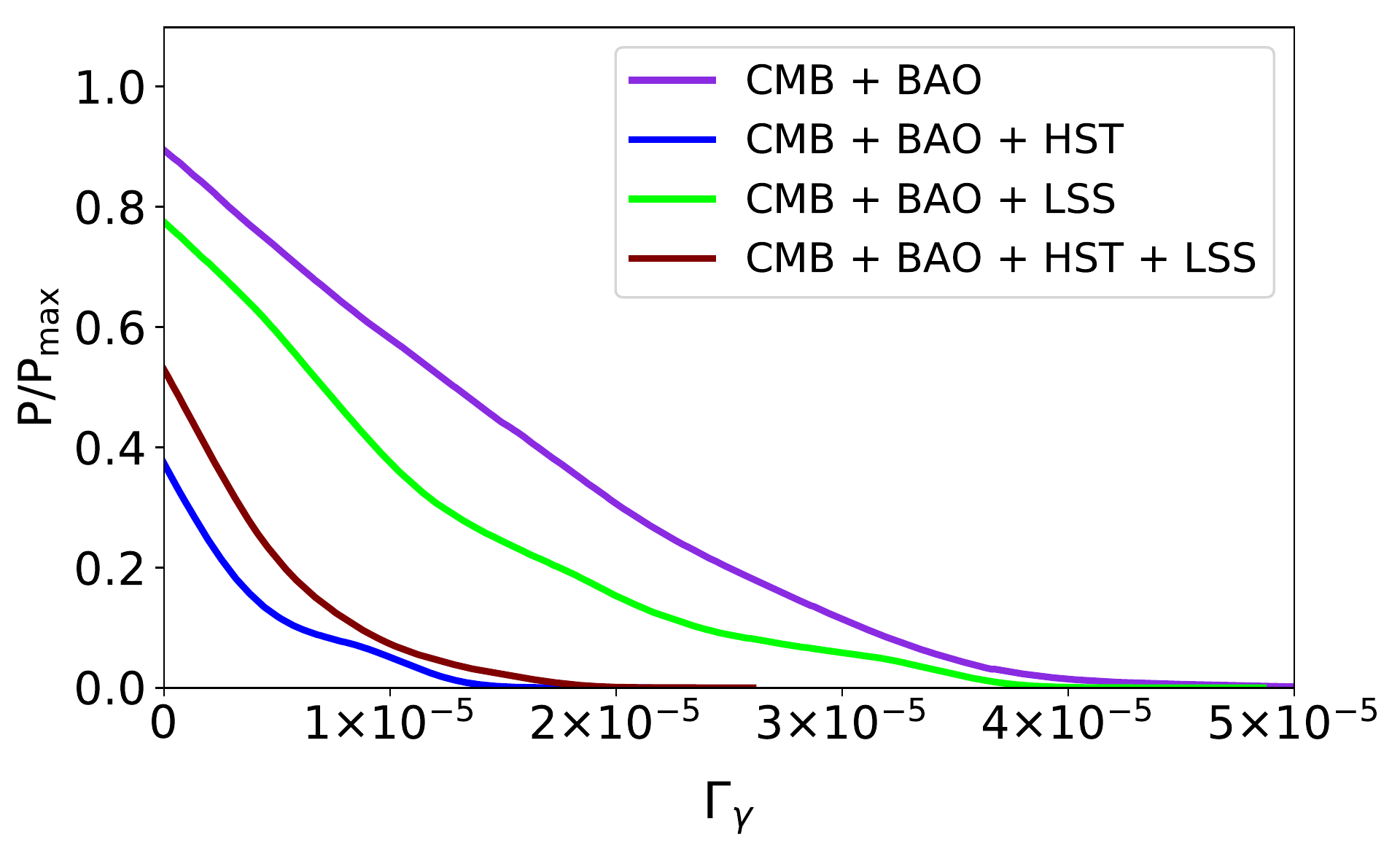}
 \caption{\label{Gamma} {One-dimensional marginalized distributions of  $w_{\rm de0}$ (left panel) and $\Gamma_{\gamma}$  (right panel).}}
\end{figure*}  
 
\begin{figure*}[!ht]
\centering
\includegraphics[width=7.5cm]{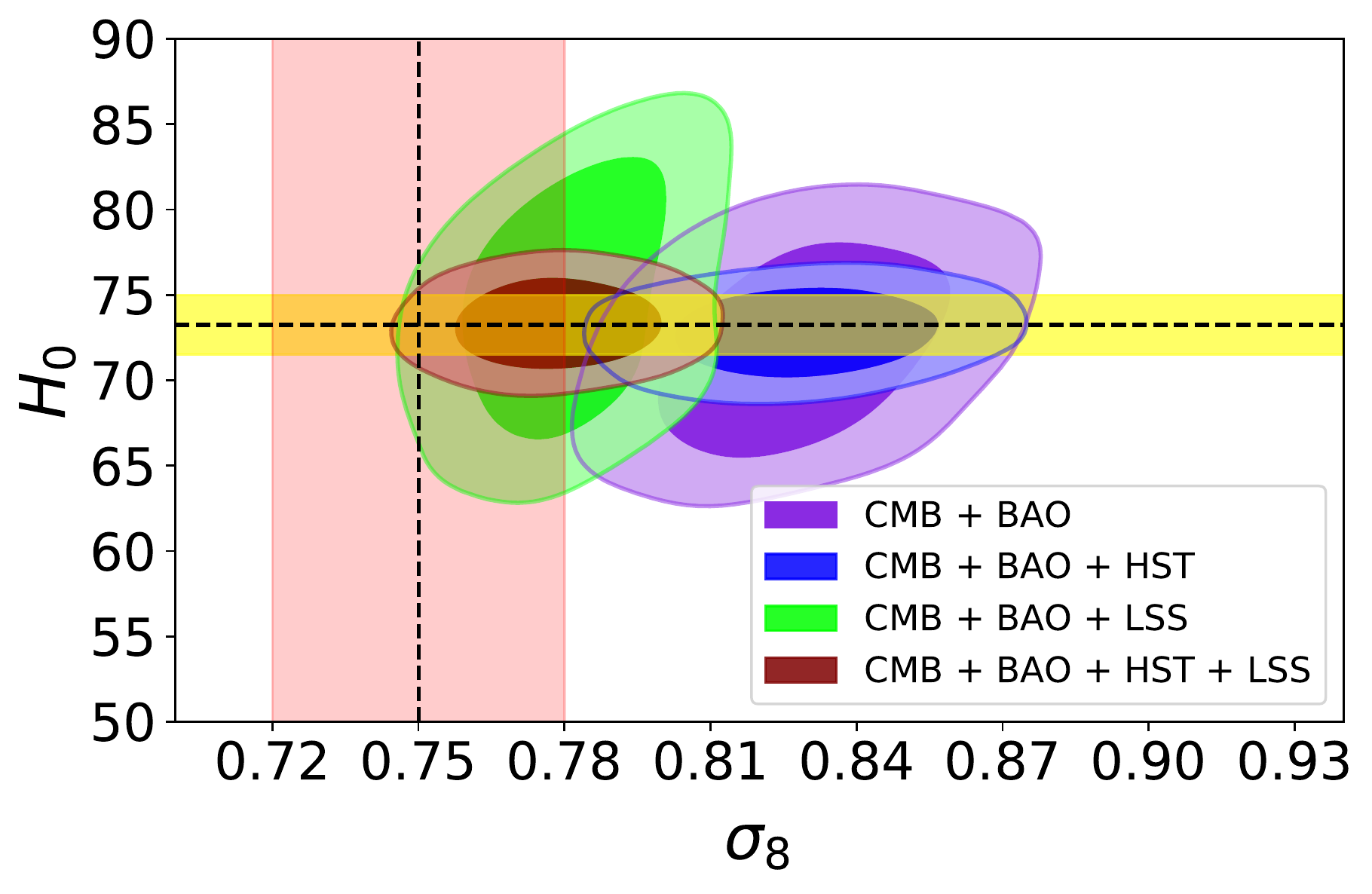}
\includegraphics[width=7.5cm]{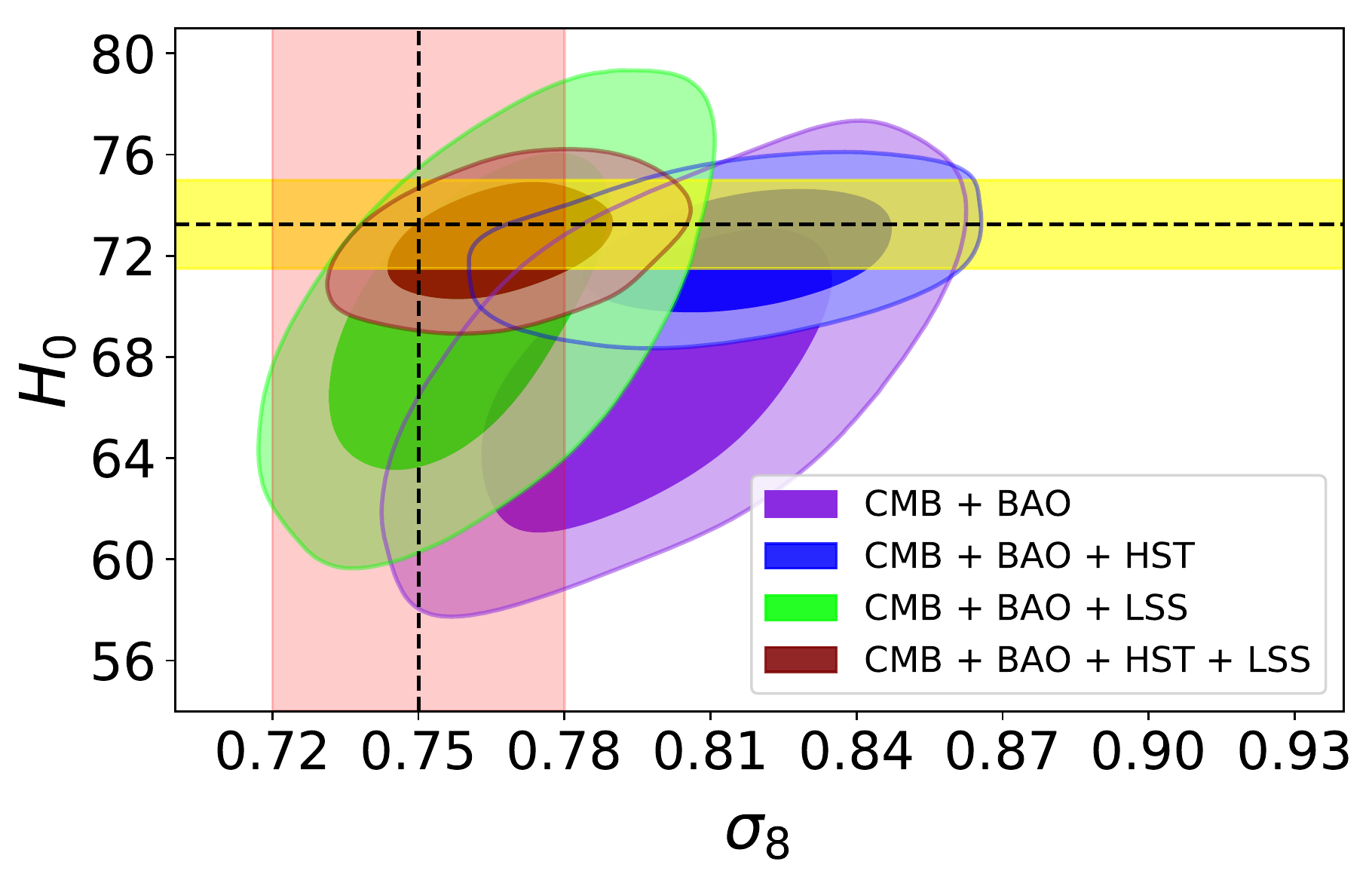}
\caption{\label{hs8} { 68\% and 95\% confidence contours  
for $H_0$ and $\sigma_8$ in our previous study \cite{DDM07} (left panel) and in present study  (right panel). In both panels, the horizontal yellow band shows local value $H_0=73.24\pm 1.74$ km s${}^{-1}$ Mpc${}^{-1}$, reported by Riess {\it et al.} \cite{riess} whereas the vertical light red band represents the Planck-SZ measurement: $\sigma_8= 0.75\pm 0.03$ \cite{LSS1}.}}
\end{figure*}

\begin{figure*}[!ht]
 \includegraphics[width=15.5cm]{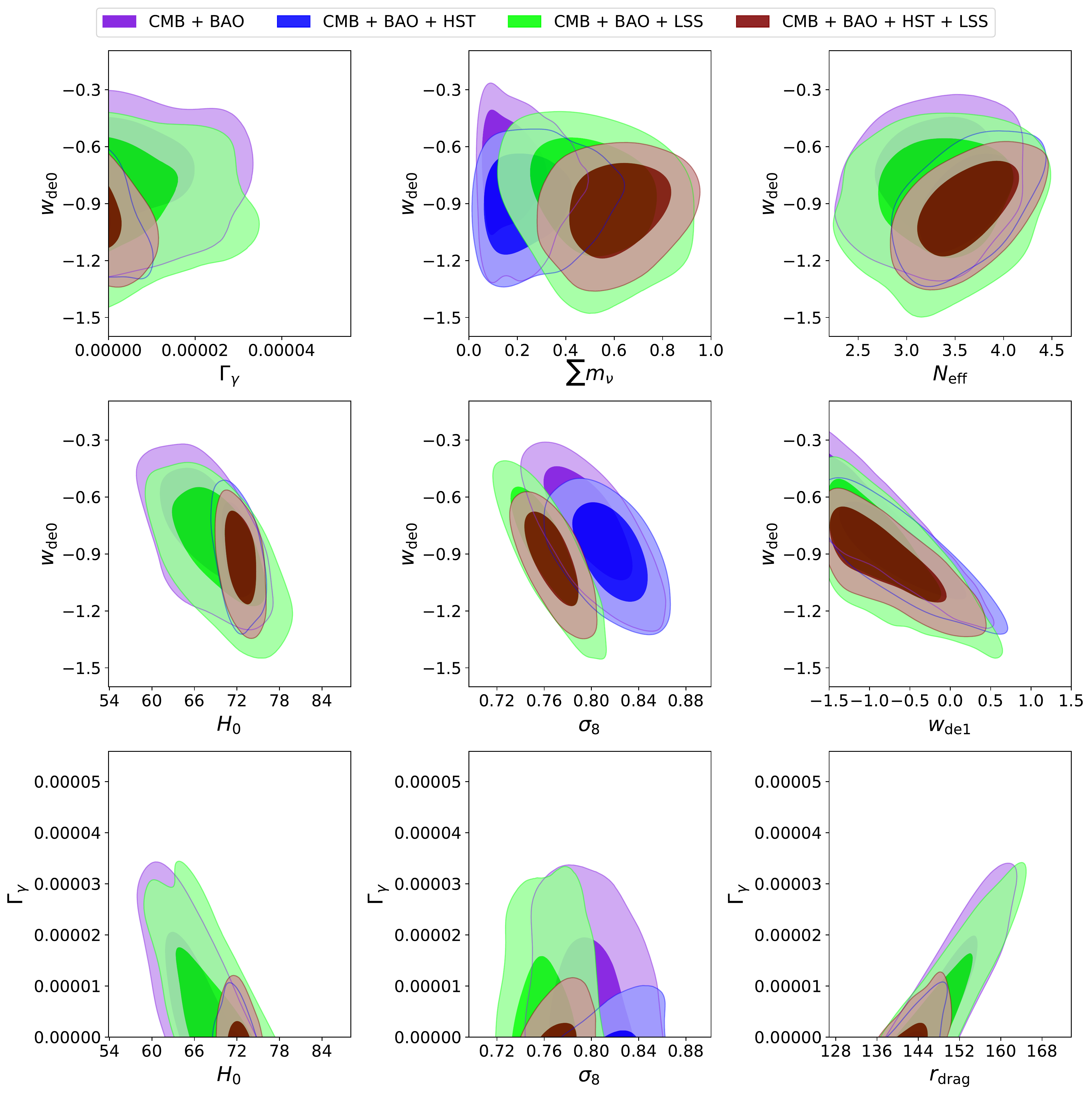}
 
\caption{\label{2dplot} {68\% and 95\% confidence contours for some selected model parameters.}}
\end{figure*}   

Next, we discuss the impact of the time-varying EoS of DE on $H_0$ and $\sigma_8$ in the context of the well-known tensions on these parameters investigated in several studies \cite{H00, H01,H02,H03,H04,H05,H06a,H06b,H06c,H07,H08,H09,H10,H11}. In Fig. \ref{hs8}, we have shown the parametric space of $H_0-\sigma_8$ obtained in our previous study (left panel) in contrast with the present study (right panel), where the horizontal yellow band shows local value $H_0=73.24\pm 1.74$ km s${}^{-1}$ Mpc${}^{-1}$, reported by Riess {\it et al.} \cite{riess} whereas the vertical light red band represents the Planck-SZ measurement: $\sigma_8= 0.75\pm 0.03$ \cite{LSS1}. We notice a clear deviations in the probability regions of the  $H_0$ and $\sigma_8$ parameters resulting due to the inception of time-varying DE in the present study. These deviations are useful to alleviate the $H_0$ and $\sigma_8$ tensions, as discussed in the following.

From Table \ref{Table_M1}, one can see that with the base data set: CMB + BAO, $H_0 = 67.4\pm 3.9$ Km\,s$^{-1}$\,Mpc$^{-1}$ at 68\% CL, consistent with the Planck measurement \cite{Planck2015}. However, with the inclusion of HST and LSS data to the base data, we have $H_0 = 72.2\pm 1.6$ Km\,s$^{-1}$\,Mpc$^{-1}$  and $H_0 = 69.8\pm 4.1$ Km\,$s^{-1}$\,Mpc$^{-1}$, both  at 68\% CL, respectively. In the joint analysis, we obtain $H_0 = 72.5\pm 1.5$ Km\,s$^{-1}$ \,Mpc$^{-1}$ at 68\% CL. Thus addition of HST and LSS data yields larger values of $H_0$, in line with the local value $H_0 = 73.24\pm 1.74$ Km\,s$^{-1}$\,Mpc$^{-1}$ as reported by Riess {\it et al.} \cite{riess}. In the present analysis, we have obtained lower mean values on $H_0$ in all the four cases as compared to our previous work \cite{DDM07} but still consistent with the local measurement at 68\% CL (see Fig. \ref{hs8}). 

With regard to $\sigma_8$ tension, one can note from the Table \ref{Table_M1} that we have obtained lower values with $\sigma_8= 0.799 ^{+0.020}_{-0.026}$, $\sigma_8= 0.761 ^{+0.017}_{-0.021}$ and $\sigma_8= 0.767 ^{+0.014}_{-0.016}$, all at 68\% CL from CMB + BAO, CMB + BAO + LSS and the joint analysis, respectively. These values are in good agreement with the  direct measurements like galaxy cluster count, weak gravitational lensing and Sunyaev-Zeldovich cluster abundance measurements, etc. However, with the case CMB + BAO + HST, we have $\sigma_8= 0.816 ^{+0.022}_{-0.020}$ at 68\% CL, favouring Planck CMB measurement. We observe that variable EoS of DE provides slightly lower values of $\sigma_8$ with all data combinations as compared to our previous results in \cite{DDM07}. In particular, a significant change is observed with the data combinations: CMB + BAO and CMB + BAO + LSS. One may see the consistency of the range of $\sigma_8$ values in the present study, with the Planck-SZ measurement $\sigma_8= 0.75\pm 0.03$ \cite{LSS1} (also see the right panel of Fig. \ref{hs8}). 

Further, one may see the correlation of present DE EoS parameter $w_{\rm de0}$  and DM-photon coupling parameter $\Gamma_\gamma$ with some other model parameters in Fig. \ref{2dplot}. In particular, we observe that $w_{\rm de0}$ shows a negative correlation with $\sigma_8$ and $H_0$ parameters with all data combinations. Thus, higher values of $w_{\rm de0} $ correspond to lower values of $\sigma_8$. In general, we notice that $w_{\rm de0} $   and $\Gamma_\gamma$ show correlation  with all other parameters especially in case of the  full data combination. Next, we have found the upper bound on the neutrino mass scale as $\sum m_\nu<0.89$ eV at 95\% CL with joint analysis: CMB + BAO + HST + LSS. We notice that the constraints on neutrino mass scale are similar to those obtained in our previous study \cite{DDM07}, with all data combinations.  Also, one can see from Fig. \ref{2dplot} that $w_{\rm de0}$ does not exhibit correlation with neutrino mass scale $\sum m_\nu$. Thus, we observe that time-varying DE does not have any significant effect on the neutrino mass scale. Next, in comparison  to our previous work, here we have found significantly lower values  $N_{\rm eff} = 3.29\pm 0.39$ and $N_{\rm eff} = 3.40\pm 0.40$, both at 68\% CL with CMB + BAO and CMB + BAO + LSS, respectively. The constraints on $N_{\rm eff}$ with other two data combinations are consistent  with our previous work. We have obtained constraints on $r_{\rm drag}$ similar to our previous work \cite{DDM07}. In Fig. \ref{2dplot}, one can see a positive correlation between $\Gamma_\gamma$ and $r_{\rm drag}$.  The constraints on $r_{\rm drag}$ are in good agreement with the recent  measures in \cite{Planck2015, rs01,rs02} at 68\% CL. 
\section{Statistical Model Comparison}

In this section, we perform statistical comparison of the considered model with a known well-fitted reference model (here we have chosen the $\Lambda$CDM model). For this purpose, we use classical statistical criterion, namely, the Akaike Information Criterion (AIC) \cite{akaike74, burnham02}, derived from information theory and defined as 
 
 \begin{equation} \nonumber
 \text{AIC} = -2 \ln  \mathcal{L}_{\rm max} + 2{\mathcal N} \quad = \chi_{\rm min}^2 + 2 \mathcal{N},
 \end{equation}
 where $ \mathcal{L}_{\rm max}$ is the maximum value of the likelihood function for the model, and $\mathcal{N}$ is the total number of estimated parameters in the model. To compare the considered model $i$ with a reference model $j$, we need to determine the difference of AIC values  of the two models, i.e.,  $\Delta{\rm AIC}_{ij} =  \text{AIC}_{i}- \text{AIC}_{j}$. This difference can be used to interpret the evidence in favor of the model $i$ compared to the model $j$. As argued in \cite{tan12}, one can confidently declare that one model is better than the other if the difference of AIC values of the two models is greater than a threshold value $\Delta_{\rm threshold}$. The thumb rule of AIC states that $\Delta_{\rm threshold} = 5$  is the universal value of threshold irrespective of the properties of the model considered for comparison. It is clearly stated in \cite{liddle07} that this threshold value is the minimum required difference of AIC values of the two models to strongly claim that one model is better in comparison to the other model. Thus, an AIC difference of 5 or more favors the model with smaller AIC value. Also, a model with large number of parameters is penalized in AIC criterion.

Table \ref{bayesian_evidence} summarizes the difference of AIC values, i.e., $\Delta \rm AIC$ of the considered model with respect to the standard $\Lambda$CDM model for all data combinations. We have found that $\Delta\rm AIC$ value is greater than the threshold value for the data combinations: CMB + BAO and CMB + BAO + HST. Therefore, it can be claimed that with these two data combinations, the standard $\Lambda$CDM model is strongly favored over the model under comsideration.  For the other combinations: CMB + BAO + LSS and CMB + BAO + HST + LSS, we can not claim statistical evidence in favor or disfavour of either of the models on the basis of AIC difference since $\Delta \rm AIC$ is less than the threshold value.  Although, a mild statistical preference of the considered model is observed in the joint analysis: CMB + BAO + HST +LSS. 
 
\begin{table}{}
\centering
\caption{\label{bayesian_evidence}{}Difference of AIC values  of considered model with respect to minimal $\Lambda$CDM model with considered data combinations. }
\begin{tabular}{l c  }
\hline \hline
Data  &  $\Delta \rm AIC$  \\
\hline
CMB + BAO      &        9.56     \\
CMB + BAO + HST  &         7.36   \\
CMB + BAO + LSS   &     -1.34    \\
CMB + BAO + HST + LSS    &     -4.80    \\

\hline \hline
\end{tabular}
 \end{table}
}

%%%%%%%%%%%%%%%%%%%%%%%%%%%%%%%%%%%%%%%%%%
\section{Final Remarks}
\label{remarks}
In this Paper, we have investigated DM-photon coupling model with time-varying EoS of DE via CPL parametrization. We have observed significant changes due to the time-varying EoS of DE on various model parameters by comparing with our previous study \cite{DDM07}, where a constant  EoS of DE was assumed. We have found that in the DM-photon coupling scenario the mean value of $w_{\rm de0}$ favors quintessence behavior ($w_{\rm de0}>-1$) of DE with all data combinations (see left panel of Fig. \ref{Gamma}). We have observed significant correlations of the DE EoS parameter $w_{\rm de0}$ with other model parameters (see Fig. \ref{2dplot}). Due to the decay of DM into photons, we have obtained higher values of $H_0$, consistent with the local measurements, similar to our previous study. In addition, the time-varying DE leads to lower  values of $\sigma_8$ in the DM-photon coupling model with all data combinations, in comparison to the results in our previous study. Thus, allowing time-varying DE in the DM-photon coupling scenario is useful to alleviate the $H_0$ and $\sigma_8$ tensions  (see Fig. \ref{hs8}).
\subsection*{Acknowledgments}
The author sincerely thanks to S. Kumar and R. C. Nunes for fruitful discussions and suggestions in improving the manuscript. The author acknowledges the Council of Scientific \& Industrial
Research (CSIR), Govt. of India, New Delhi, for awarding Senior Research Fellowship (File No. 09/719(0073)/2016-EMR-I).

\end{document}